# *FarView*: An In-Situ Manufactured Lunar Far Side Radio Array Concept for 21-cm Dark Ages Cosmology

Ronald S. Polidan[a,1], Jack O. Burns[b,a], Alex Ignatiev[a], Alex Hegedus[c], Jonathan Pober[d], Nivedita Mahesh[e], Tzu-Ching Chang[f], Gregg Hallinan[e], Yuhong Ning[d], Judd Bowman[g]

[a]Lunar Resources, Inc., 18108 Point Lookout Drive, Houston, TX 77058 USA
[b]Center for Astrophysics and Space Astronomy, University of Colorado Boulder, 593 UCB, Boulder, CO 80309, USA
[c]SRI International, 2100 Commonwealth Avenue, Ann Arbor, MI 48105 USA
[d]Department of Physics, Brown University, Box 1843, Providence, RI 02912 USA
[e]Cahill Center for Astronomy and Astrophysics, California Institute of Technology, MS 249-17, Pasadena, CA 91125, USA
[f]Jet Propulsion Laboratory, California Institute of Technology, MS 138-308, 4800 Oak Grove Drive, Pasadena, CA 91109, USA
[g]School of Earth and Space Exploration, Arizona State University, ISTB4-675, Mail Code 6004, Tempe, AZ 85287

**Abstract**
*FarView* is an early-stage concept for a large, low-frequency radio observatory, manufactured in-situ on the lunar far side using metals extracted from the lunar regolith. It consists of 100,000 dipole antennas in compact subarrays distributed over a large area but with empty space between subarrays in a core-halo structure. *FarView* covers a total area of ~200 km$^2$, has a dense core within the inner ~36 km$^2$, and a ~power-law falloff of antenna density out to ~14 km from the center. With this design, it is relatively easy to identify multiple viable build sites on the lunar far side. The science case for *FarView* emphasizes the unique capabilities to probe the unexplored Cosmic Dark Ages – identified by the 2020 Astrophysics Decadal Survey as the discovery area for cosmology. *FarView* will deliver power spectra and tomographic maps tracing the evolution of the Universe from before the birth of the first stars to the beginning of Cosmic Dawn, and potentially provide unique insights into dark matter, early dark energy, neutrino masses, and the physics of inflation. What makes *FarView* feasible and affordable in the timeframe of the 2030s is that it is manufactured in-situ, utilizing space industrial technologies. This in-situ manufacturing architecture utilizes Earth-built equipment that is transported to the lunar surface to extract metals from the regolith and will use those metals to manufacture most of the array components: dipole antennas, power lines, and silicon solar cell power systems. This approach also enables a long functional lifetime, by permitting servicing and repair of the observatory. The full 100,000 dipole *FarView* observatory will take 4 – 8 years to build, depending on the realized performance of the manufacturing elements and the lunar delivery scenario.

---

[1] Corresponding author at: Lunar Resources, Inc., 18108 Point Lookout Drive, Houston, TX 77058 USA. *E-mail address:* ron@lunarresources.space (R. S. Polidan).

**Introduction**

*FarView* is an early-stage concept for a low frequency (5-50 MHz) radio telescope array comprised of 100,000 dipole antennas, distributed over ~200 km$^2$ on the lunar far side, which could be built in the 2030s. It is designed to study the Universe's "Dark Ages," the period of cosmic history before the ignition of the first stars. *FarView's* ultimate deliverable will be a tomographic map tracing the evolution of the Universe across a huge swath of cosmic history stretching from before the birth of the first stars through the early Cosmic Dawn epoch. Decameter emission from the cosmic Dark Ages cannot penetrate the Earth's ionosphere, and anthropogenic radio frequency interference (RFI) will compromise these observations from being made on Earth or in Earth orbit, including the Lagrange Points. In the inner solar system, the only place these observations can be conducted with sufficient sensitivity is on the Moon's far side, and even there one must avoid areas near the lunar limb because refracted RF noise from the Earth will impact sensitivity. With 100,000 in-situ built antennas, *FarView* will be one of humanities' next "great" observatories – the most sensitive, powerful, low frequency radio telescope ever built. *FarView* will be capable of uncovering revolutionary new insights into the formation of the first structures in the Universe, dark matter, dark energy, and cosmic inflation, thereby dramatically enhancing the standard models of physics and cosmology.

What enables this observatory and makes it feasible and affordable in the timeframe of the 2030s is that it is manufactured in-situ, utilizing space industrial technologies developed by Lunar Resources, Inc. (LUNAR): Molten Regolith Electrolysis (MRE) hardware that extracts metals (e.g., Fe, Si, Al, and Mg) and oxygen from lunar regolith; Space Deposition System (SDS) which uses the extracted metals to vapor deposit antennas and solar cells on the lunar surface; and Pulsed Electrical Discharge (PE3D) technology that uses the extracted metals in additive manufacturing of required components for the observatory. These in-situ resource utilization (ISRU) technologies can produce nearly all components needed to build *FarView*. Dipole antennas and power lines are comprised of thick film metallic strips deposited directly on the lunar surface by manufacturing rovers. The antennas are then integrated with preamps (Earth-built or partially in-situ built) for processing and transmission of the data. Power will be supplied by thin film silicon solar cell panels, also fabricated in-situ by a separate facility, and transported by rovers to the build locations for integration into the power grid. The full *FarView* observatory will take 4 – 8 years to build, depending on the realized metal extraction rate, antenna manufacturing rate, rover performance, and hardware delivery scenarios. However, *FarView* will be built in a subarray configuration under which each new subarray will be activated upon completion and start generating data in concert with other completed subarrays, thus supplying scientific data from the beginning of construction. *FarView* can accommodate almost any delivery scenario and can utilize large or medium landers to deliver needed hardware to the Moon. Building *FarView* will also enable major new lunar developments by demonstrating large scale and sustained lunar ISRU.

Building *FarView* on the lunar far side requires identifying locations that are sufficiently flat to permit manufacturing and that are also far enough from the lunar limb (≥25° Bassett et al 2020) to have low RFI levels (see also Le Conte, Elvis, & Gläser 2023). The concept for the *FarView* array architecture is adapted from the large, Earth-based, low frequency radio arrays and consists of compact antenna subarrays widely dispersed in a core-halo arrangement, across a large area of the Moon, with each subarray containing 300-600 dipoles, laid out in rows of antennas. Each subarray has a small footprint (~0.15 km$^2$), including power and communications infrastructure, thus facilitating finding suitable areas for the manufacturing process. The number of dipoles within a subarray is somewhat arbitrary and is determined by topography, manufacturing, and computational constraints, rather than science requirements. This array layout is optimized for analysis of the Dark Ages, with adjacent dipoles oriented orthogonally and separated by ~17 m. This observatory distribution (~6 km core, ~ 14 km halo, orthogonal adjacent dipoles) yields measurements of all four Stokes parameters (albeit non-co-spatial) and facilitates removal of foreground signals. *FarView* will operate during lunar days (e.g., solar observations) and nights (e.g., Dark Ages observations), imaging the entire sky above the lunar horizon with each integration. Since each subarray is independently powered, science data acquisition will start as soon as the first subarrays are built and powered up; there is no need to wait until the full complement of dipoles are built.

*FarView* builds upon earlier lunar radio telescope concepts including Takahashi (2003), Jester & Falcke (2009), Boonstra et al. (2010), Lazio et al. (2011), Mimoun et al. (2012), Klein-Wolt et al. (2012), Zarka et al. (2012), Bentum et al. (2020), Koopmans et al. (2021), Burns et al. (2019, 2021b), and Chen et al. (2024).

This paper first presents the scientific motivation for building the *FarView* radio observatory on the lunar far side, and the importance of Dark Ages hydrogen cosmology to our understanding of the origin and structure of the Universe. The *FarView* team has completed a feasibility study and is beginning a design study that will define the system architecture, establish performance goals, develop delivery and build timelines, and mature the technical details of the observatory. Details and results from our initial science simulations appear in Section 2. These include a discussion of the array configuration forecasting, array sensitivity, approach to foreground mitigation, and expectations for *FarView's* performance. Section 3 then presents preliminary results on how *FarView* will be implemented, describing the technologies that enable *FarView* to be constructed on the lunar surface with lunar materials, followed by a discussion of the overall observatory architecture and physical infrastructure (antennas, electronics, power, rovers, and communications). The final portion of this section provides a top-level overview of how *FarView* will be physically constructed and operated. Section 4 summarizes the paper's key findings and discusses possible next steps.

1. *FarView* Science, Design, and Expected Performance

    2.1 Scientific Motivation: Dark Ages Hydrogen Cosmology

The 21-cm line of neutral hydrogen (HI) offers the only known observational probe of the Universe's "Dark Ages," the period of cosmic history before the formation of the first stars; however, this technique requires observations at wavelengths longer than that allowed by Earth's ionosphere, necessitating a mission in space. A series of U.S. National Academy and Science Community Reports have recognized the potential for breakthrough discoveries of "HI Cosmology," especially from the far side of the Moon. NASA's *Enduring Quests, Daring Visions* Astrophysics Roadmap (Kouveliotou et al. 2013), and the *Artemis III Science Definition Team Report* (2020) both emphasize the great promise of these Dark Ages observations at low radio frequencies using a lunar far side interferometric array. The *Pathways to Discovery in Astronomy and Astrophysics for the 2020s* Decadal Survey report singled out the Dark Ages as THE Discovery area in cosmology for this decade. Here, we develop the core science requirements for *FarView* as well as the construction and data-processing designs for such a Dark Ages experiment.

Neutral hydrogen can be detected through its hyperfine 21-cm transition (referring here to the rest wavelength of the emitted line = $\lambda_{rest}$). The observed wavelength of the line ($\lambda_{obs}$) grows significantly through cosmological redshifting ($\lambda_{obs} = (1+z)\, \lambda_{rest}$, where $z$ is the redshift): 21-cm emission from the cosmic Dark Ages ($z \gtrsim 30$) is expected at decameter wavelengths or frequencies ($\nu_{obs} = c/\lambda_{obs}$) ~5-50 MHz today. By measuring the signature over a wide range of wavelengths and angular scales, one enables a three-dimensional reconstruction of the distribution of matter (since redshift can be mapped to distance). The ultimate deliverable of HI cosmology is therefore a tomographic map tracing the evolution of the Universe across a huge swath of cosmic history stretching from before the birth of the first luminous objects through the early Cosmic Dawn (first stars) epoch. We note here that the Dark Ages and the earliest portion of the Cosmic Dawn are only accessible from the Moon's far side, whereas the later Epoch of Reionization (EoR) may be probed using ground-based arrays such as LOFAR[2], MWA[3], HERA[4], and SKA[5] at higher frequencies ($\gtrsim 50$ MHz).

Figure 1 places the Dark Ages, Cosmic Dawn, and EoR into perspective. After the Big Bang, the Universe was hot, dense, and nearly homogeneous. As the Universe expanded, the material cooled, condensing after ~400,000 years (z~1100) into neutral atoms, freeing the Cosmic Microwave Background (CMB) photons. The baryonic content during this pre-stellar Dark Ages of the Universe consisted primarily of neutral hydrogen. About fifty million years later, gravity propelled the formation of the first luminous objects – stars, black holes, and galaxies – which ended the Dark Ages and commenced the Cosmic Dawn (e.g., Loeb and Furlanetto 2013). These first stars (Pop III, nearly metal-free, ~100 solar mass) likely differed dramatically from stars we see nearby, as they formed in vastly different environments (Abel et al. 2002). Figure 2 illustrates the evolution of structures in the early universe as viewed by the redshifted 21-cm signal along with the "global" or all-sky spectrum and the amplitude of spatial fluctuations for several structure modes (see below).

The first data released from the *James Webb Space Telescope* (*JWST*) point (albeit provisionally) toward important new insights into early galaxy formation (e.g., Castellano et al. 2022, Donnan et al. 2022, Mason et al. 2023, Mirocha & Furlanetto 2023). However, *JWST* will likely only scratch the surface of the late Cosmic Dawn because the sources are so faint and distant (Behroozi & Silk 2015), and it will be unable to probe the Dark Ages era targeted by *FarView* because there are no galaxies. Other near-future facilities, like *SPHEREx*[6] and HERA, will also study the last phases of Cosmic Dawn. We can leverage these complementary probes to improve our models of the 21-cm spin-flip signal and optimize the design of lunar radio telescopes (e.g., Dorigo Jones et al. 2023).

Spatial fluctuations in the 21-cm Dark Ages signal are governed almost entirely by well-understood linear structure formation, the same physics used to interpret *Planck* and other observations of the CMB power spectrum, allowing precise predictions of the expected signal within the standard cosmological model. Interferometric measurements of fluctuations in the 21-cm Dark Ages signal can therefore uniquely test the standard cosmological model at the onset of structure formation, without the complication of highly non-linear baryonic effects (Chen,

---

[2] https://www.astron.nl/telescopes/lofar/
[3] https://www.mwatelescope.org/
[4] https://reionization.org/.
[5] https://www.skao.int/en
[6] https://www.jpl.nasa.gov/missions/spherex.

Meerburg, and Münchmeyer 2016, Loeb and Zaldarriaga 2004). Any departure from these well-constrained predictions will provide important new insights into the physics of structure formation, potentially into the nature of dark matter (Slatyer et al. 2013), early dark energy (Hill & Baxter 2018), or any exotic physics (Clark et al. 2018). Fundamentally, such observations could also measure the ultimate number of linear modes (independent Fourier modes of the 3D density field) in the Universe and lead to exquisite cosmological constraints, including the masses of neutrinos and their hierarchy (Mao et al. 2008), the non-Gaussianity of initial density perturbations (Muñoz et al. 2015a, Cooray 2006, Mondal & Barkana 2023, Bull et al. 2024), and the imprints of primordial gravitational waves to reveal the complexity and energy scale of cosmic inflation (e.g., Book et al. 2012, Schmidt et al. 2014, Ansari et al. 2018, Furlanetto et al. 2019).

In many ways, spatial fluctuations in the 21-cm absorption during the cosmic Dark Ages provide the ultimate cosmological observable. The simplest way to quantify these fluctuations is with the power spectrum (Figure 3), which characterizes the amplitude of the variations as a function of spatial scale ($1/k$, where $k$ is the wavenumber), analogous to *Planck* measurements of the CMB. During this time, the 21-cm line traces the cosmic density field with most modes in the linear or mildly non-linear regime, allowing a straightforward interpretation of the measurement in terms of the fundamental parameters of our Universe (Lewis & Challinor 2007). The lack of luminous astrophysical sources makes the Dark Ages signal a clean and powerful cosmological probe and renders the 21-cm line the *only* known observable signal from this era.

Furthermore, the 21-cm line can be used to reconstruct a 3D volume (as compared to the 2D surface of the CMB) and is not affected by "Silk damping" (e.g., Hu, Sugiyama, & Silk 1997) on the smallest scales, which blurs out fluctuations in the CMB — meaning the number of accessible modes is enormous and is effectively only limited by the collecting area of the instrument. (Silk damping is also known as photon diffusion damping which reduces density anisotropies in the early Universe, making both matter and the CMB more uniform.) Because 21-cm measurements can access an enormous number of modes in the linear regime inaccessible by other means, they enable stringent constraints on the physics of the early Universe, and particularly its early phase of inflation.

The 21-cm power spectrum is also an exquisite probe of physics occurring during the Dark Ages themselves. All the physical processes that affect the 21-cm all-sky or global signal (see e.g., Burns et al. 2021a), including the exotic scenarios (e.g., nongravitational cooling of HI via scattering by partially charged dark matter) proposed for explaining the EDGES signal (Bowman et al. 2018), also affect the power spectrum. But the power spectrum has far more information so allows more precise tests of the scenarios. We note that the power spectrum can only be measured with an interferometer, where a large range of angular scales can be sampled which are inaccessible to a large single dish telescope. FarView can also measure the Dark Ages global spectrum (i.e., the monopole) by using one of the subarray outriggers along with the halo antennas to measure the foreground.

Moreover, many exotic processes imprint distinct signatures in the Dark Ages power spectrum (see e.g., Figure 3 and Muñoz et al. 2018). Dark matter remains a mystery and the cosmic Dark Ages offer the best astrophysical probe of dark matter physics on cosmic scales. For example, any warm dark matter (e.g., Hibbard et al. 2022 and references therein) will suppress the formation of small structures and hence the amplitude of the power spectrum on those scales. Constraining such scenarios with observations targeting later epochs is complicated by the slew of baryonic feedback processes that also operate on small scales. As a result, redshifted HI observations targeting epochs after the Cosmic Dawn suffer a degeneracy between astrophysical processes that halt star formation in low-mass dark matter halos and cosmological processes that suppress the formation of those dark matter halos. Observations aimed at the Dark Ages avoid this important degeneracy.

### 2.2 Array Configuration Forecasting for *FarView*

Decameter emission from the cosmic Dark Ages is distorted and absorbed by the Earth's ionosphere and limited by radio frequency interference, so must be observed from space[7]. However, even from the far side of the Moon, the HI signal from the Dark Ages will be exceptionally challenging to measure. At all redshifts, the HI signal is buried underneath orders of magnitude of foreground emission: radio emission from the Milky Way and other galaxies that obscure the cosmological HI. Foreground emission is dominated by synchrotron radiation which steeply increases in brightness towards longer wavelengths. The strength of the HI signal also evolves with redshift, but in a complex way that traces the various radiation backgrounds permeating the Universe (see Figure 2 and Mesinger et al. 2014). The effect is something of a coincidence: at redshifts ~2, ~8, and ~20, the foreground-to-signal ratio is a fairly consistent $10^{4-5}$ (Pritchard and Loeb, 2008). However, during the cosmic Dark Ages, the foregrounds have grown substantially brighter while the HI signal strength in the standard cosmology model has dropped – making the expected foreground-to-signal ratio more like $10^{5-6}$ at the end of the Dark Ages ($z$~30). Removing foreground emission from Dark Ages HI cosmology space missions will therefore be even more challenging than for ground-based experiments (but like that needed for CMB B-mode experiments). Thus, there are two key questions that need to be considered when forecasting the required array configuration for an experiment like *FarView*: (1) Will *FarView* possess the intrinsic sensitivity needed to detect the faint HI signal? (2) Will *FarView* be able to remove or otherwise mitigate foregrounds to the degree necessary to recover the HI signal?

The Dark Ages signal strength is expected to be very faint. This signal has yet to be detected, but expectations are based upon robust predictions from current models within the standard physics/cosmology paradigms; new physics likely enhances the signal strength. In addition, there are uncertainties in the lunar environment and the performance of radio antennas on the lunar surface. Using the best available Dark Ages models and incorporating the expected performance of dipole antennas on the lunar surface following the analysis by Koopmans et al. (2021), we expect a minimum of 50,000 dipoles will be needed to detect the signal. Given this uncertainty, to provide adequate science margin (sensitivity scales as square root of the number of dipoles), we scaled this number up by a factor of two, to 100,000 dipoles. Improved, model calculations and data from mid-decade precursor missions such as *ROLSES*, *LuSEE-Night*, and *FARSIDE*[8] will remove many of these uncertainties and the *FarView* architecture, specifically the number of dipoles required will be revisited. One major advantage of *FarView* is that because it is built with in-situ materials, the number of dipoles in the array can be adjusted as data are gathered from the first subarrays that are built and brought into operation.

---

[7] During night-time conditions at solar minimum, the ionosphere's plasma cut-off frequency can fall to 10-20 MHz ($\lambda$=15-30 meters), allowing observations to high redshifts; however, nights of such quality are few, and recent studies suggest ionospheric scintillation can corrupt precision measurements at frequencies as high as 50-100 MHz (e.g., Vedantham and Koopmans, 2015, Datta et al. 2016, Shen et al. 2021). Furthermore, human generated radio interference is a major source of contamination for long-wavelength radio experiments. An array on the far side of the Moon will be uniquely shielded from this interference; this isolation has been another major driver for a lunar radio telescope.

[8] See https://www.colorado.edu/ness/ness-projects for a summary of each project. ROLSES and LuSEE are single antenna experiments currently manifested on CLPS/PRISM missions to the Moon. FARSIDE is a candidate Probe-class mission proposed to the Astro2020 Decadal Survey, an array of 128 pairs of dipole antennas.

So, the basic *FarView* design that was developed in our initial engineering study – 100,000 dipole antennas, split between a dense central core and "halo" of outrigger" stations – is motivated by the need to answer both the above questions. The dense core provides the surface brightness sensitivity necessary to detect a faint, diffuse HI background signal. The antennas far from the core provide higher angular resolution necessary to precisely model the foreground emission.

**2.3 Sensitivity Forecasting**

To answer question (1) – *Will FarView possess the intrinsic sensitivity needed to detect the ultra-faint HI signal?* – we begin by recognizing that the sensitivity is determined by the number and type of antennas used in the array. We assume 10-m tip-to-tip thin-wire antennas and a total of 100,000 antennas. With an array diameter of ~14 km, *FarView* achieves a resolution of ~7 arcmin at 15 MHz, which is sufficient to resolve cosmological structures at scales within the highest signal/noise regime (see Figure 8 in Koopmans et al. 2021). The array sensitivity depends upon the frequency bandwidth, the integration time, the system temperature, and the effective area of the array. The effective collecting area for each antenna is determined by the dipole length and the antenna impedance, and this was modeled using NEC4.2 simulations which consider that the dipole wires rest directly on the regolith (~30 m$^2$ per dipole or ~3 km$^2$ for full array at 30 MHz). The system temperature depends upon the sky and regolith temperatures, and the front-end amplifier. The array RMS sensitivity (ΔS), or minimum detectable flux density, is given by

$$\Delta S = \frac{2kT_{sys}}{A_{dipole}\sqrt{N(N-1)(\Delta f)(\Delta t)}}, \quad (1)$$

where $A_{dipole}$ is the effective area of one of the dipole antennas, N is the total number of dipoles in the array, $\Delta f$ is the frequency bandwidth, and $\Delta t$ is the integration time. The dipole antennas are electrically short yet are efficient enough to be sky-noise dominated (see Burns et al. 2019 for details on modeling of the antennas on the lunar surface). So, the system temperature ($T_{sys}$) is given by the Galaxy brightness temperature (produced by synchrotron emission above ~5 MHz),

$$T_{sys} \sim 5000 \left[\frac{\nu}{50MHz}\right]^{-2.5} \text{ K}. \quad (2)$$

After only 1 minute of integration with $\Delta f$=0.5$f$ and with $f$ as the center frequency of the band, the RMS sensitivity for broadband total power measurements, as would be used to measure, e.g., solar and exoplanetary system radio bursts, is ~2 mJy at 15 MHz and ~0.2 mJy at 40 MHz.

For 21-cm cosmology measurements, we follow the basic analysis given by Koopmans et al. (2015), used for the design of the Square Kilometer Array (SKA), to estimate the power spectrum error due to thermal noise. This is given by

$$\Delta^2_{noise} \propto \left(\frac{A_{core}}{\sqrt{N}A_{coll}^{3/2}}\right) \quad (3)$$

where *N* is the number of antennas, and $A_{coll}$ is the effective collecting area of the core of the array (collecting area of each dipole times [*N(N-1)*]$^{1/2}$/2). It is assumed that the antennas are distributed over a tightly packed core area, $A_{core}$, so that the density in the u-v (projected baseline) plane is sufficiently dense. From this equation, the power spectrum sensitivity (at a given *k*-mode scale) is driven by (a) the total collecting area, and (b) the compactness of the array. The sensitivity also depends weakly on the field-of-view.

A more rigorous approach to determine an optimal distribution of antennas for *FarView* requires numerical simulations. *FarView* presents a novel challenge compared with existing and upcoming arrays: its large number of antennas. The key to an HI cosmology experiment's sensitivity is its "u-v coverage," i.e., the footprint of its sampling pattern in the Fourier plane projected onto the sky. Each pair of antennas (or "baseline") in an interferometer samples a unique Fourier mode of the sky (see Figure 4). We use the code *21cmSense* (Pober et al. 2016) which begins its sensitivity forecast by calculating all the Fourier modes sampled by an instrument, including the time-dependence of this sampling (since a given point on the sky moves over the course of an observation). Because the critical ingredients to this calculation are the unique separation vectors between every pair of antennas in the array, the computation grows quadratically with the number of antennas. Therefore, compared with current and proposed arrays featuring several hundred antennas, *FarView* is in an entirely different regime. Preliminary results with a modified (but very approximate) *21cmSense* indicate that 0.3 km$^2$ of collecting area in the dense core for *FarView* will be sufficient for a 3$\sigma$ detection of the HI signal from the end of the Dark Ages at z~30 over a total observation period of ~5 years. More collecting area will likely be required for earlier times (higher redshifts, lower frequencies) in the Dark Ages as the foregrounds grow much brighter thus increasing the thermal noise. These requirements are consistent with the analysis by Koopmans et al. (2021) who calculate that ~1 km$^2$ of collecting area can deliver a ~10 $\sigma$ detection at z~ 30 also over 5 years integration. Full *21cmSense* forecasting, which is planned in the next phase of our *FarView* study, will provide us with tighter constraints on the instrument design that can deliver its science goals. It is important to note that since we are building *FarView* using lunar regolith source material, it is possible to add more dipoles or modify the shape to fit the results from the next phase of this study.

### 2.4 Foreground Mitigation

At the scale of *FarView*, answering our second question – "Will *FarView* be able to remove or otherwise mitigate foregrounds to the degree necessary to recover the HI signal?" – is a more daunting proposition. The last decade has seen several ground-based experiments including the LOw-Frequency ARray (LOFAR; van Haarlem et al. 2013), the Murchison Widefield Array (MWA; Tingay et al. 2013, Wayth et al. 2018), and the Hydrogen Epoch of Reionization Array (HERA; DeBoer et al. 2017) – place increasingly stringent limits on the HI signal strength at frequencies above ~100 MHz (corresponding to the EoR epoch), but there has been no successful detection. The limiting factor for all these experiments is their ability to remove foregrounds in the presence of a complicated, time- and frequency-dependent instrument response. Foreground removal is therefore, in many ways, a work in progress.

These experiments have taught us many lessons about how the design of an instrument (and, in particular, its spectral response) complicates the removal of foregrounds from the data. Effects once deemed negligible for radio interferometry have been identified as critical points of the analysis to model and understand. The discussion in the previous section in and around equation (3) suggests that splitting *FarView* 50-50, between a compact core to measure the 21-cm power spectrum and an extended halo of longer baselines to measure the foreground, is a reasonable estimate. We propose to precisely characterize the foreground using the longer baselines in the outer array halo for mitigation/subtraction using techniques such as delay spectrum analysis mentioned below. It is worth noting that due to the high-z/low-frequency nature of the Dark Ages measurement, the foreground avoidance technique common in the ground-based 21-cm cosmology field will not be applicable for *FarView*, due to the limited 21-cm window in the cylindrical wavenumber *k* space.

The field of 21-cm cosmology has now reached a point where high-precision instrument simulations (i.e., those that create reliable mock data) can be used to evaluate both instrument designs and analysis pipelines. At the forefront of this effort is *pyuvsim*, an ultra-precise simulator for radio interferometers, developed by J. Pober and collaborators (Lanman et al. 2019). As already discussed, the *FarView* design features significantly more antennas than current instruments. In moving forward, we plan to use *pyuvsim* to evaluate several key performance metrics for the *FarView* antenna design. In particular, the single-baseline "delay spectrum" (Parsons et al. 2012, Lanman et al. 2020) has become a powerful tool for HI cosmology; using the delay spectrum applied to single-baseline *pyuvsim* simulations, we can evaluate the frequency response of the *FarView* antenna design and ensure that it introduces no spectral structure beyond what is already tolerated for ground-based instruments. Furthermore, several interferometric data simulation packages are under construction that employ approximations which speed up the calculations. A campaign is underway in the HI cosmology community to reference these simulators to *pyuvsim* and to see what approximations introduce acceptable levels of errors.

### 2.5 Beamforming Subarrays

Performing interferometry for each pair of antennas in the *FarView* array would require $N^2$ ~$10^{10}$ correlations. As noted above, this exceeds our current capability to numerically simulate the design and construction of *FarView* by combining all the signals in the array from individual dipoles. An alternative is "beamforming" that effectively synthesizes a single aperture and beam, which in this case is applied to each subarray. The phasing of each subarray can produce a desirable beam and pointing of the subarray. Effectively, each subarray can then be treated as an antenna for the purpose of modelling the optimization of *FarView* for 21-cm cosmology. Ground-based arrays such as the SKA will take advantage of beamforming for their operations.

A trade study was performed on beamforming subarrays versus full interferometry for *FarView*. The reference subarray design is 300-500 dipoles, each spaced by ≈17 meters using a non-cospatial layout design. This size is based on the number of antennas that can be deployed by ISRU/vacuum vapor deposition in a single lunar day; it is not based on science estimates and can be updated. We also note that the fringe rate is 28 times slower on the Moon compared to Earth. The results of the trade study are shown in Table 1. The table lists criteria for performance evaluation, our relative weights, and our scores for beamforming subarrays versus interferometry of all the array dipoles. This analysis suggests a slight preference for beamforming although it might result in a loss of some science. Others may provide different weights and scoring. We also recognize that the reduced field of view in beamforming will diminish the array sensitivity so the details of the beamforming approach will be revisited as we conduct more detailed sensitivity studies using numerical simulations. In moving forward with the detailed design of *FarView*, we will assume subarray beamforming although this could change in future years as technologies advance.

### 2.6 Simulations of *FarView*'s Performance

We used an existing framework to simulate the approximate performance of the configuration and point spread function of the *FarView* array. This framework was introduced in Hegedus et al. (2020) and used again in Burns et al. (2021b) for simulating the *FARSIDE* array. This simulation pipeline utilizes a combination of the Common Astronomy Software Applications (*CASA*) radio astronomy software package (McMullin et al. 2017), lunar surface maps from *LRO LOLA* (Barker et al. 2016), and JPL space-mission observation geometry package *SPICE* (Acton 1996) to define a simulated array on the Moon's surface. This approach properly aligns the reference frames from the lunar surface to the sky for the purpose of observing low frequency emissions. Because of the computational load, we modelled the performance assuming beamforming of the subarrays which we then treat as effective antennas.

Accurate estimates of the predicted brightness maps along with realistic noise profiles must be used to ensure that a given array size/configuration is sufficient to observe the target to a certain noise level or resolution. The geometry of the array must also be known to a high degree so that the baselines are computed accurately to enable correct imaging of the data. Currently, the best maps of the lunar surface are Digital Elevation Models (DEMs) from *LRO/LOLA*. The simulation pipeline accepts longitude and latitude coordinates for each receiver and interpolates the elevation at these points from existing DEMs to calculate the full 3D position. The DEM used here has a resolution of 128 pixels per degree of longitude at the lunar equator. We use the configuration of the *FarView* array with a core-halo distribution and combined with a DEM around the lunar Pauli impact basin as shown in Figure 5.

From these coordinates, the baselines are computed and inserted into a *CASA* Measurement Set (MS) file. The MS format can be used with many existing analysis and imaging algorithms, and holds information about the array configuration, visibility data, and alignment with the sky. However, many of these software routines are hard coded to work with arrays that rotate with the Earth's surface, and do not work for orbiting or lunar arrays. To circumvent this, this simulation pipeline manually calculates the baselines and visibilities for a given array and imaging target and inserts the data into the MS format. The *SPICE* library is used to correctly align the lunar and sky coordinate systems and track the motions of the Moon, Earth, and Sun. For these early simulations, visibilities of a target image were not computed. The u-v-w (3D projected baselines onto the sky) column of the MS that defines the antenna separations was filled out for each pair of *FarView* subarrays, of which there are $160 \times 159/2 = 12720$ pairs.

These data were then imaged and "cleaned" (i.e., deconvolved) using the widefield imaging software *WSClean* (Offringa et al 2014). Running this software on the generated *FarView* MS observation file yields a reconstructed image of the sky brightness, as well as the point spread function (PSF) of the array. This imaging currently uses a uniform weighting of the baselines, so if multiple visibilities fall in one u-v cell, they are down-weighted by the number of visibilities in that cell. In this weighting mode, the image will have the highest resolution, but the system noise will not be optimal. The output point spread functions (PSFs) from this program were then analyzed to quantify the performance of the *FarView* array. Figure 4 shows the results from an extended integration of the sky over 5 Earth days (0.2 lunar days) to increase the quality of the reconstruction. The panels in Figure 4 show the reconstructed point spread function, the u-v coverage of the observation, a histogram of the baseline lengths, and statistics on the quality of the PSF as a function of distance off zenith, with lower sidelobes leading to a better reconstruction. This set of figures assumed an idealized beam pattern and that each subarray has the same configuration and response. Additionally, no frequency chromaticity effects were included here since the figure shown is a PSF of only a single frequency.

In summary, our preliminary analysis indicates that *FarView* will need to have a combination of a compact core with a ~6 km diameter, that we estimate to contain a uniform spacing of about half the 100,000 antennas for enhanced sensitivity to the Dark Ages power spectrum, and an approximate power-law distribution of the other half of the antennas from the core to an outer diameter of ~14 km, to precisely measure the foreground. For construction purposes, we will build antennas in subarrays with ~300-500 antennas/subarray. With the configuration shown in Figure 4, *FarView* will be able to provide unprecedented data and insight into the earliest stages of the formation and evolution of structure in the Universe and discover potential new physics during the Dark Ages. Further analyses must be performed to develop key design elements regarding the characteristics of the array needed to achieve the science goals. These new analyses tools combined with the adaptability of the in-situ manufacturing approach will, with high confidence, ensure that *FarView* will be able to deliver these most challenging science goals. Because *FarView* is built from lunar materials on the Moon's surface, the array architecture is adaptable as we evaluate its performance with real data. This means that *FarView* can evolve and be upgraded to meet evolving science goals much as is done with current Earth-based radio interferometers.

## 3.0 Mission Implementation

*FarView* is a science-driven mission. It is enabled by advanced ISRU technologies, specifically the extraction of metals from lunar regolith and the use of those metals to manufacture almost all the observatory components directly upon the lunar surface. Hence, the architecture and implementation of *FarView* must be a product of the combined requirements for the science and manufacturing. It must also fit within programmatic constraints, such as a reasonable lunar delivery schedule, being built in an acceptable amount of time, having a long operational lifetime, and having an acceptable cost/risk profile.

In this section we briefly discuss the enabling technologies and then present an overview of the *FarView* observatory architecture, implementation approach, and science operations plans.

### 3.1 Enabling Technologies

The fundamental enabling component of *FarView* is large scale lunar ISRU. It would not be possible to build *FarView* if all observatory components had to be transported from Earth – the cost and risk would be prohibitive. However, by landing ~5000 kg of space industrial equipment (excluding rovers and communication hardware) to extract ~10-15 tonnes of refined metals from the lunar regolith per year and using these metals to manufacture antennas and almost all the supporting infrastructure, *FarView* becomes economically feasible with a greatly reduced cost and risk relative to alternative approaches. The in-situ manufacturing of functional components on the lunar surface is facilitated by the extensive work done by LUNAR in ISRU technology maturation.

Three of these ISRU technologies are essential to enabling *FarView*: 1) Molten Regolith Electrolysis, 2) Space Vacuum Vapor Deposition, and 3) Metals Additive Manufacturing.

### 3.1.1 Molten Regolith Electrolysis

The extraction of metals from the lunar regolith that are essential for *FarView* fabrication (aluminum, magnesium, silicon, and iron) requires the reduction of the metallic oxides that comprise the regolith. There have been multiple approaches on regolith reduction, but they have principally focused on NASA's interest in the extraction of oxygen, with the majority of the approaches requiring some reagent/consumable brought from Earth. *FarView* will use an electrolytic approach for regolith reduction, thus requiring only electrical energy to extract metals. The electrolytic process requires melting the regolith to achieve electrical conductivity/electrolysis and is thus called Molten Regolith Electrolysis (MRE). MRE was initially developed in Sadoway's group at MIT (Sadoway, et al 2008, Sirk et al 2010, Vai et al 2010) and by Schreiner (Schreiner et al 2015). This technology has been transferred to LUNAR for metals extraction. It is currently at TRL-4, will be at TRL-5 by mid-2024, and with current funding should reach TRL-6 by 2025. MRE's sister technology, Molten Oxide Electrolysis has also been commercialized (Allanore 2014) for steel production. MRE involves applying a potential and creating a current between an anode and a cathode immersed in a molten regolith magma which reduces the oxides, generating oxygen and metal atoms based upon the Gibbs Free Energy of Oxide Formation. The basic reductions reactions are:

$$2Al_2O_3 \rightarrow 3O_2 + 4Al$$

$$SiO_2 \rightarrow O_2 + Si$$

The extracted aluminum is used to manufacture dipoles and power lines and the extracted silicon is used to manufacture solar cells.

As noted, the regolith must be molten to conduct electrolysis current which requires MRE operation at temperatures exceeding 1600 C. LUNAR has developed the high temperature reactors required for MRE and has devised a two-step process for extraction of high purity aluminum that can then be utilized for antenna fabrication. Laboratory experiments have successfully demonstrated the extraction of aluminum, silicon, iron, and oxygen from regolith simulant.

### 3.1.2 Space Vacuum Vapor Deposition

For *FarView*, the high purity metals extracted by MRE, specifically aluminum, need to be fabricated into metallic dipole antennas and power lines. This is accomplished through application of the space vacuum vapor deposition technology which was originally developed and demonstrated under NASA's Wake Shield Facility program (3 Shuttle flights, 1994-96; Ignatiev 1995), and will be used to fabricate thin aluminum strip lines in the vacuum environment of the Moon by LUNAR's Space Deposition System (SDS). The SDS technology is currently at TRL-5 and will attain TRL-6 prior to its demonstration mission on a space flight planned for 2025. The *FarView* utilization of SDS will require an upgrade of the existing hardware for higher flux deposition as the *FarView* dipoles require a metallic thickness of at least the extinction distance of the detected radio waves (~100 microns). This thick aluminum film is fabricated by vapor deposition from an electrically heated effusion cell. Both the rate and throw of the cell can be controlled to specifically deposit a 100-micron thick, 3 cm wide aluminum strip line on lunar regolith. A prototype strip line has been successfully fabricated in a vacuum chamber on a lunar regolith simulant. In addition to antenna fabrication, the SDS will also be used for fabrication of the power lines to each antenna preamp as well as for the feed power lines from the lunar power source (see Section 3.4). The SDS unit is relatively small and can be mounted almost anywhere on the rover that allows it access to the lunar surface. It will require between 300 W to 600 W during the deposition of the antennas and the power lines which is within the projected power capacities of possible *FarView* rovers. It is currently envisioned that rovers will be re-supplied by raw materials by returning to the MRE base but could also be re-supplied by a small transport rover delivering raw materials to the rover site.

In addition to the dipole antennas, and the interconnecting power lines, the SDS tool can be used to fabricate various thin film device structures in the vacuum of the Moon using the aluminum and the other extracted elements like iron and silicon. One such important thin film device using silicon and aluminum is the thin film silicon solar cell. Such a solar cell fabricated in-situ on the surface of the Moon can be used to supply the power needed not only for the fabrication of the *FarView* antennas, but also for the power needed to operate *FarView*. The thin film silicon solar cells are fabricated on molten regolith glass substrates and are microcrystalline in nature. As a result, their efficiencies are lower than terrestrial silicon cells: however, the in-situ process as well as the very large available area within a subarray (100's m$^2$) allows for the fabrication of large numbers of cells thereby giving the ability to supply almost any amount of energy needed for operation on the Moon. Such thin film silicon solar cells have been successfully fabricated in a terrestrial vacuum chamber. On the Moon the solar cells would be robotically fabricated near the materials extraction area using a modified SDS system capable of depositing interleaved layers of silicon and aluminum to form the thin film silicon cell. The projected production rate for the in-situ process is near ~100 kW per lunar day of solar panels which will have an estimated functional life span of ~10 years.

### 3.1.3 Metal Additive Manufacturing

The availability of metals from the MRE also drives the possibility of additive manufacturing of components using those metals. Some of the challenges in the additive manufacturing of metals on the Moon include restriction on the use of binder agents, both in transport costs and vacuum non-compatibility, and the high power required for the metal melting operation (laser, microwave, e-beam). To mitigate these challenges, the Pulsed Electrical Discharge additive manufacturing (PE3D) technology uses a pulsed power process for metal melting. PE3D can be used to melt metals such as aluminum and iron that will be available from MRE on the Moon, as well as refractory metals such as niobium, molybdenum, and others. Due to the use of pulsed power and due to a unique electronics design, the PE3D process requires less than 500 W of supplied power to melt metals, which is within the range of payload power supported by the rovers being considered for use by *FarView*. For *FarView*, PE3D will be utilized for fabricating the interconnects required for the powerlines and antennas and will also be evaluated for antenna fabrication as compared to the SDS vapor deposition process. The PE3D would be mounted on a robotic arm on a rover so that the printing capabilities can be manoeuvred to any point on the surface to print interconnects or antennas. This location on the robotic arm also allows PE3D to be utilized for repair of power lines or antennas which have been physically damaged and require re-connection.

These ISRU technologies currently range from TRL-3 to TRL-5 and are actively under development external to *FarView* for application in much larger projects with earlier delivery dates. Their current development schedule has them reaching TRL-6 within two years, with anticipated deployment of prototypes to the lunar surface by 2028. Laboratory experiments with these technologies have extracted metals from regolith simulant, have vacuum deposited thick metallic films, and have printed metallic structures, and have fabricated thin films solar cells. These are the necessary building blocks that need to be in place for the 100,000-antenna *FarView* observatory to be built. The *FarView* concept is sufficiently adaptable that it should allow use of these ISRU technologies with only minor modifications to their primary application design. Utilization of these technologies for in-situ fabrication also enables another critical capability for *FarView*: servicing and repair. As components degrade in the lunar environment, the assets (materials and hardware) are already in place to fabricate new units or repair degraded ones. *FarView* can be maintained at a full science performance level for decades with only occasional replacement of hardware from Earth and could last 50+ years.

### 3.2 *FarView* Architecture

The science requirements presented in §2 provide a sufficient level of definition, when combined with the ISRU requisites, to establish a reasonable observatory architecture. As discussed above, the overarching configuration is a core-halo distribution of subarrays with half the subarrays in the ~6 km diameter core and the remaining subarrays in a quasi-power law falling off to 14 km from the center of the observatory.

Science performance also sets requirements for the location of the observatory. As stated in §2, the location must be in the far side radio quiet zone. It also must be relatively flat and relatively free of physical hazards like rocks, craters, and ridges. (Note that all sites will require some level of surface preparation before construction can commence.) It should ideally be free of magnetic anomalies and, ideally, the site should also have a clear horizon and a locally uniform, thick, subsurface crust. An additional site selection factor will be the surface aluminum abundance. For this initial assessment of the surface abundance of metals we used lunar surface composition measurements from Zhang et al. (2023) and Srivastava, (private communications (2022, 2023). *FarView* needs to be built in a region with a higher surface aluminum abundance (~50-70 MT will be needed to build the observatory). This requirement will eliminate a few sites, but a range of abundances can be accommodated by varying the size of the extraction area (a very manageable ~300 m$^3$ volume of regolith is needed for an Al composition of 10% by mass).

Accommodating all these requirements clearly will constrain the suitable sites, but examination of lunar maps and lunar surface physical attributes has yielded multiple sites that appear to meet all the required site criteria. Table 2 lists 8 locations that appear to meet all site selection criteria (First Tier) and 8 additional locations that meet most of the site selection criteria (Second Tier) along with their selection criteria shortfalls. During the design study these sites will be explored in detail, and the list of potential build locations will be reduced to the two or three best sites. Science performance requirements set the dipole antenna size, separation, and distribution within a subarray, but the number of dipoles in the subarray is only loosely constrained by science performance. The number of dipoles in a subarray is primarily determined by the rate at which the dipoles are manufactured by the SDS hardware, and the computational load needed to acquire and process the signal. Current estimates for this dipole production rate are between 300 and 600 dipoles per lunar day. With dipole production only occurring during lunar days, it is most efficient to complete production of one subarray per lunar day per manufacturing rover. Given this rate range and the desire to complete one subarray per rover per day, we have chosen a 400-dipole subarray as our baseline subarray size. Figure 6 illustrates the layout of a baseline subarray that meets all the science and manufacturing constraints. This subarray has a small footprint (~0.15 km$^2$ – roughly a ~385 m square), facilitating the identification of locations that can accommodate ~250 subarrays in a core-halo distribution.

In summary, addressing these location requirements will reduce selection, but overall, it is anticipated that multiple fully suitable sites will be identified as the analysis progresses.

### 3.3 Antenna Preamplifiers

The preamplifiers needed for the dipole antennas present a significant issue for *FarView*. While it is relatively straight-forward to manufacture the preamplifiers on Earth and transport them to the Moon, transporting 100,000 items, even if they are of individual small mass, will be expensive. As such, two technical approaches are being pursued: developing ultra-low mass Earth-built preamplifiers; and developing a hybrid preamplifier in which most of the unit is built in-situ using the thin film deposit process used in microelectronics technology with minimal Earth-built components. To mitigate the effects of the harsh lunar environment the preamplifier modules will be buried 30 to 40 cm below the lunar surface with cabling connecting it to the surface antennas and power lines. At these depths, the temperature is nearly constant year-round near the average temperature at the location (Figure 7 and see Xiao et al. 2022 and Feng and Siegler 2021). For likely *FarView* locations this will be around -80 C. The density of the regolith at these shallow depths is small ($\leq$ ~1.2 g/cm$^2$, see Xiao et al. 2022) so burying the units should not present major difficulties. The intervening regolith will also provide some protection from micrometeoroids and radiation. If needed, a small amount of heater power can be added to maintain the desired temperature for the electronics. Application of more cold tolerant electronics for the receiver/preamps are being explored in the ongoing design study to reduce the need for heaters.

### 3.4 Power Infrastructure

Two separate power networks will be implemented for *FarView*. The site where we extract and process the metals needed for manufacturing the observatory, manufacture the solar arrays, perform data processing, and have the Moon-Earth communication hardware will have a dedicated power network. We estimate that it will need around 60 kW of power. Although this level of power can be supplied by an Earth-delivered power units, we propose the use of the in-situ fabricated photovoltaic (PV) solar arrays. These in-situ solar cells can be created at a rate of ~100 kW per lunar day once the manufacturing system is fully operational. However, starting the solar cell production process will likely require a small Earth-built power unit. Defining this startup power unit and the overall power system initiation process will be part of the NIAC Phase II analyses. The harsh lunar environment will impact the entire power infrastructure. For the solar arrays this can be mitigated by the fabrication of new solar cells; however, studies are currently underway exploring cell efficiency in the lunar radiation/thermal/dust environment with the goal of identifying characteristics that make them more tolerant to the lunar environment. Earth-manufactured power generation technologies that could supply that level of power nominally with multiple units are in the early stages of development for the *Artemis* Program but were not evaluated during the feasibility study. These will be evaluated during the design study.

The second power network is the one that powers the subarrays. Each subarray is <u>individually</u> powered (Figure 6), rather than in a single interconnected power system for the whole observatory. The subarray power is supplied by the in-situ built solar panels which can be easily scaled to fit the need. Subarray sizes are kept small, and power cable lengths minimized to reduce power loss. We estimate that each subarray will need around 1 W per dipole antenna, plus additional power for data processing and communications. For a 400-dipole antenna subarray this equates to ~500 W of power. This need is far below the estimated solar cell production rate of 100kW per lunar day even if future analyses indicate a lower solar cell production rate or lower cell efficiencies or greater power needs for the subarrays.

The in-situ fabricated solar arrays will supply the lunar day *FarView* power needs. For nighttime, however, energy storage will be required. It is well to note that optimal operation of the *FarView* observatory will be during the lunar night to eliminate solar interference, hence the power requirement for nighttime operation. This power need may be supplied by a variety of technologies including batteries. Of particular interest is using in-situ manufactured batteries, which may be available by the 2030s.

As mentioned above, the preliminary analyses of the nighttime power needs yielded an estimate of around 1 W per antenna, but with considerable uncertainty. In particular, the power needs for the acquisition and processing of the data and heaters were only notionally included. For 400 antenna subarrays this would require an energy storage system capable of supplying at least 400-500 W of continuous power for 14 days of darkness. The ongoing design study is looking in detail at the preamp/receiver design, data acquisition/processing, and advanced electronics technology to determine nighttime power needs, and the likely energy storage technologies to address those needs. While there are no readily available systems capable of supplying this nighttime power, multiple national and international government agencies, aerospace industries, and research institutions are developing concepts for providing power during lunar nights. Many of these nascent technologies (e.g., in-situ manufactured batteries: O'Meara and DeMattia 2021, Maurel et al 2023; and other, more exotic, technologies e.g., Regenerative Fuel Cells, Pu et al 2021), if matured, seem capable of supplying all *FarView's* nighttime power needs.

### 3.5 Lunar Rovers

*FarView* will require multiple lunar rovers to move the MREs from the lander to a nearby extraction site, excavate regolith for ingest by the MREs, manufacture the solar arrays, transport materials from the extraction site to the subarray build-sites, build the subarrays, and many more functions. Multiple rover concepts of all sizes (e.g., the Lockheed Lunar Mobility Vehicle[9] or Astrobotic's Polaris Rover[10]) that are currently in various stages of development have been explored for application to *FarView*; many have capabilities that appear to be within the range needed to build *FarView*. Further development of both the rover concepts and *FarView* are required before the number, size distribution, and functional capabilities of the rovers can be determined. As with the ISRU technologies, the development of these rover concepts is external to the *FarView* effort. It is anticipated that the flexibility and adaptability of the *FarView* concept will allow use of these rovers with only minor modifications.

### 3.6 Data Rate and Communications

With 100,000 dipoles imaging the sky every few minutes, *FarView* will generate a prodigious amount of data. We can estimate the expected data rates using the parameters of the planned array layout and an initial correlator scheme. *FarView,* with 100,000 elements located within subarrays, will operate over a bandwidth of 50 MHz between 5 and 50 MHz, with the largest baselines in the core and halo of 8.5 km (6√2) km and 19.9 (14√2) km, respectively. To ensure reasonable coherence on the longest baselines, we will need about ~5 kHz spectral resolution, which gives roughly 10,000 spectral channels over the 50 MHz bandwidth. We must consider a correlator to accommodate the number of visibilities the radio array would produce. For *FarView*, beamforming subarrays are more efficient (see Section 2.5 and Table 1). The scaling of the computational costs is reduced from $10^{10}$ to $N_{beams}*N_{stations}*N_{antennas/stations} + N_{beams}*N_{stations}*(N_{stations}-1)/2$. The term $N_{beams}*N_{stations}*N_{antennas/stations}$ accounts for the voltages/data from all the antennas within the station that go into the beamformer. Thus, the amount of data that will be downlinked to the Earth will only depend on $N_{stations}*(N_{stations}-1)/2$, i.e., 160*(159)/2 assuming just one beam per station. Having calculated the number of frequency channels and the number of visibilities, the last factor we need is accumulation time. Given the slow rotation of the Moon, the sky above the array changes at the rate of 0.5 arcmin/min. The resolution corresponding to the longest baselines in the halo and core of the array at the center of the band (15 MHz) is ~3.4' and 8', respectively. Assuming the sky does not change within $1/10^{th}$ of the PSF, we can integrate for ~1 min and 2.4 mins for the halo and core longest baselines, leading to 1440 and 600 integrations every 24 hours. Furthermore, we assume 64-bits to store the complex visibilities (similar to the OVRO-LWA).

---
[9] https://www.lockheedmartin.com/en-us/products/lunar-mobility-vehicle.html
[10] https://www.astrobotic.com/lunar-delivery/rovers/polaris-rover/

With all these parameters, the data rate per day to be downlinked to Earth is:
- Data-rate/$24^h$ (halo) = $N_{stations}(N_{stations}-1)/2 * N_{channels} * N_{bits(ADC)} * N_{pol} * N_{integrations}$
  $= 12,720 * 10,000 * 64 * 4 * 1440$
  $= 5.8$ TB/day
- Data-rate/$24^h$ (core) $= 12,720 * 10,000 * 64 * 4 * 600$
  $= 2.4$ TB/day

We are considering alternative correlator schemes as follows to manage this high data volume:

1. **EPIC** (E-Field Parallel Imaging Correlator; Thyagarajan et al. 2017) has been successfully evaluated for ground-based low frequency interferometers such as the New Mexico LWA (Krishnan et al. 2023). It reduces the number of computations in a typical FX correlator from $N^2$ to $(N)\log(N)$. We plan to evaluate if this advantage will translate for a subarray/beamforming system.

2. **Radio Camera**. The DSA-2000 radio array[11] is replacing the traditional correlator backend with a "radio camera", an approach that may have potential for *FarView*. For DSA-2000, visibilities are not stored but rather cross-correlation, flagging, calibration, and imaging are done on a single computing platform (GPU). The array configuration (which may require modifications from the *FarView* planned redundant scheme) needs to produce an ideal PSF with minimal side lobes. This will enable fast (or remove the need for) deconvolution and directly produce images of the sky to be downlinked for every integration/accumulation time. So, the data rate will be 1440*size/image. Once again, we plan to examine how this on-the-fly calibration and imaging system will work for 21-cm cosmology.

The manufacturing hardware are relatively autonomous, so expectations for command-and-control communications during the build phase are expected to be well within expectations for lunar infrastructure in the 2030s, and continuous contact is not required. Similarly, the command and control need for the science operations phase are expected to be well within lunar communications network expectations. The preliminary *FarView* design assumes that all communications between subarrays could be performed by wireless systems. Within a subarray both wireless and wired (using deposited wires) communication systems are feasible, with the wireless system presenting fewer problems. However, establishing a wireless system in the radio quiet region of the lunar far side is problematic. While we can operate this wireless system at frequencies that do not impact *FarView* science, we want to ensure that RFI is not generated which impacts other radio science operating at other frequencies on the lunar far side. Our surface communications analysis will also explore optical communications solutions for *FarView's* surface communications needs. *FarView* will abide by all recommendations that are being developed to keep the lunar far side radio quiet. Details regarding the communications system will be developed during the ongoing design study. Even with optimistic estimates of data processing/compression, the science data volume will greatly exceed the capacity of the anticipated lunar infrastructure. Lunar communication satellites are already in orbit and in development[12] that will greatly benefit *FarView* as the first subarrays come online, but; the data volume of the full observatory will likely require a dedicated optical communications link to return the science data back to Earth. The data acquisition rate, possible data compression techniques, and the capabilities of potential lunar optical links were not sufficiently defined during the feasibility study to develop a quantitative communication architecture. These parameters will be matured during the design study and a quantitative communication infrastructure developed.

---
[11] https://www.deepsynoptic.org/overview.

### 3.7 *FarView* Implementation

*FarView* will be built robotically, with limited human involvement. Astronaut support is not required but could be very helpful during the initial set up and, occasionally, over the life of the observatory. The top-level implementation plan is to first land and deploy the extraction and manufacturing hardware, rovers, and supporting hardware on the lunar surface. This can be done in a single delivery with a large commercial lander, or in multiple deliveries with medium sized landers; *FarView* can accommodate almost any delivery scenario. A list of what products need to be delivered to the lunar surface for *FarView* was only notionally addressed during the feasibility study. Delivery of ISRU hardware, rovers, communication hardware, preamplifiers and other electronics will be required, but the size and quantity require a much more mature concept than that developed in the feasibility study. This list of required Earth-built hardware will be a major product of the design study. In addition, the design study will explore in-situ manufacture of some of the electronics using the solar panel and battery manufacturing technology. The first three mission lunar days will be spent deploying hardware, setting up power and other infrastructure for the extraction site, and starting metal processing. Since significant power is required for all ISRU activities, it is restricted to operate only during lunar day time. Beginning with the fourth lunar day, subarrays will begin construction and *FarView* operations will follow a routine:

> <u>During lunar days</u> build materials will be extracted, refined, and delivered to manufacturing rovers to construct the subarrays (dipoles and power lines), solar arrays, and other infrastructure. The subarray manufacturing process is straight-forward and the same for each subarray. A manufacturing rover is supplied with aluminum and drives to the build site and begins depositing the dipole antennas and power lines according to the desired layout. As each dipole is being built, a preamplifier is attached to the two segments of the antenna. Once the subarray is fully populated with antennas, the power infrastructure is set up and the antenna rows are connected to the power system, powered up, and performs a system check out. Any connection issues are identified and repaired, and the subarray is declared operational.

> <u>During lunar nights</u>, metal extraction and manufacturing will cease, ISRU hardware and rovers will hibernate; newly built subarrays will be powered up and system checkouts will be performed; and previously built subarrays will perform daytime and night-time science

---

[12] Such a relay satellite, called *Queqiao*, is currently in operation to support the Chinese National Space Administration's Chang'e 4 mission on the Moon's far side. ESA is building the Lunar Pathfinder communications satellite that will fly with and support the *LuSEE-Night* cosmology mission (see Section 4.0) on the far side.

observations.

*FarView* will produce valuable science as soon as the first subarray is built; it does not need to wait until the full complement of dipoles are built. This implementation approach will achieve a 100,000-antenna *FarView* within 4 to 8 years, depending upon the details and timeline for the landing of hardware, the realized extraction and manufacturing rates, and the performance characteristics of the rovers.

The in-situ approach also opens the door for near real time adaptation of the array, adjustments can be made to the antennas, subarrays, and overall architecture as the observatory is being built, should the early science data or manufacturing performance indicate areas for design improvement. Similarly, the in-situ-build approach means that *FarView* is fully serviceable: antennas and the power infrastructure can be repaired if degradation or failures occur, and new subarrays can be added to the array if necessary. This serviceability will enable *FarView* to be scientifically productive for decades with only occasional replenishment of materials from Earth. Once operational, the observatory architecture can be evolved to improve the primary *FarView* science or to enable additional, community-driven science. *FarView's* regolith processing will also provide tonnes of residual oxygen and metals as a by-product that would be available to the greater lunar economy.

The expected impact of *FarView's* ISRU on the lunar environment will be minimal. The 200 $km^2$ area of observatory will certainly alter the surface. But the regolith extraction area and the solar array manufacturing site are much smaller (a few hundred square meters) than the observatory. Of much bigger concern is developing techniques for RF shielding all the electronics (ISRU hardware, rovers, communications systems, etc.) associated with the observatory to ensure the observatory area maintains the extremely low RF background required for the science observations. In addition, there is concern regarding unintended RF emissions leaking from satellites orbiting the Moon. Unintended low frequency RF signals have been detected from Earth orbiting satellites by LOFAR (Di Vruno et al. 2023). Methods to eliminate RF interference from observatory electronics will be addressed in the design study.

It is premature to present detailed costing of *FarView*. A notional cost estimate for *FarView* was made during the feasibility study but with so many key components of *FarView* at low TRL, it should not be considered reliable. A preliminary cost estimate will be developed in the design study as the concept and its required technologies mature, but this too will have a large uncertainty. The capabilities and costs associated with the delivery of hardware to the Moon and for lunar rovers and communications infrastructure in the 2030s is particularly uncertain and will likely remain so at the end of the 2-year design study. The feasibility study estimate of the ISRU hardware was in the $1-2 B range with considerable uncertainty given their current TRL status. This cost estimate does not include any R&D development costs for the ISRU hardware and lunar rovers because these developments are part of much larger, non-astronomical lunar programs, unrelated to *FarView*. The *FarView* program plan is to purchase additional units of these components from the primary customer, perform any minor modifications needed, and deliver them to the Moon for use.

During the feasibility study we also examined the cost for a fully Earth-built *FarView*. The key differences would be replacing the few tons of ISRU hardware with 50 to 70 tonnes of antennas and power cabling and replacing the in-situ manufactured solar arrays with Earth-built power systems. Both approaches (in-situ and Earth-built) will need the same rovers, communication systems, receiver/preamplifiers, data processing systems, and other infrastructure. In addition, a major negative for an Earth-built system will be its inability to service, repair, or rebuild failed systems. To achieve the long productive science lifetime of the in-situ *FarView* would require frequent resupply deliveries from Earth, greatly increasing the cost.

### 3.8 Additional Implications

Validation of this ISRU system will advance lunar utilization well beyond the *FarView* mission. This system can be landed anywhere on the Moon to support a wide range of applications – it only needs power and a robotic support infrastructure to produce the metals and oxygen. This ISRU system will enable the building of a highly capable infrastructure almost anywhere on the Moon by providing lunar facilities with three of the most valuable commodities: power, oxygen, and raw metals.

## 4.0 Summary and Next Steps

We have presented the results of a feasibility study for a large, low frequency radio telescope concept, *FarView*, that could be built on the far side of the Moon in the 2030s. The key science objective of this observatory will be to explore the Cosmic Dark Ages and the beginnings of the Cosmic Dawn. This observatory has the potential to provide unique insights into dark matter, early dark energy, neutrino masses, and the physics of inflation. The high sensitivity needed to perform these measurements is supplied by the enormous size of the observatory: 100,000 ten-meter dipoles distributed across an area of ~ 200 km$^2$. The full observatory can achieve a broadband sensitivity level of 2 mJy at 15 MHz and 0.2 mJy at 40 MHz for a 1-minute integration. Since *FarView* is an interferometer, it can sample a large range of angular scales and measure the redshifted 21-cm power spectrum, which is inaccessible to a large single dish telescope such as the LCRT concept (Bandyopadhyay et al. 2021). In addition to early universe science, *FarView* can make unique contributions to solar physics, exoplanet studies, and radio transients. During lunar nights, *FarView* will focus upon 21-cm Cosmology measurements and during lunar days it will address solar and other science measurements.

What makes *FarView* achievable and affordable in the 2030s timeframe is that it is manufactured in-situ, utilizing space industrial technologies. This in-situ manufacturing architecture utilizes Earth-built hardware that is transported to the lunar surface to extract metals from the regolith and will use those metals to manufacture all 100,000 dipoles and the power infrastructure (solar arrays, batteries, and power lines). Extraction hardware takes in lunar regolith and outputs refined metals which are then used to manufacture the observatory dipoles and power infrastructure directly on the lunar surface by utilizing lunar rovers, outfitted with manufacturing hardware. The observatory architecture is a core-halo distribution of subarrays, not markedly different from large terrestrial radio arrays.

The in-situ manufacturing approach also enables near real time servicing and evolution of the array. Adjustments can be made to any aspect of the observatory as it is being built or after the observatory is fully operational. The observatory can be serviced (repairing or replacing failed components) and evolved (adding more subarrays), yielding a very long-lived facility that can adapt to new science discoveries.

The immediate next step for *FarView* is to develop a more comprehensive and quantitative system architecture. This will be done via a two-year funded design study (NIAC Phase II) that will focus upon maturing the science requirements through detailed modelling, performing vacuum chamber tests of multiple technologies to deposit antenna test articles on regolith simulant, and further advancing the system implementation from its current notional concept into a full-fledged mission implementation plan.

In parallel, NASA's commercial lunar payload services (CLPS) missions are providing the first "ground truth" for operating radio telescopes on the lunar surface. This includes the *ROLSES* radio science telescope that landed at the Malapert A crater region near the Moon's south pole on 22 February 2024 using the Intuitive Machines Odysseus lander (Burns et al. 2021b), and an upgraded *ROLSES-2* on another CLPS lander to be deployed on the lunar near side in 2026. In addition, *LuSEE-Night* will be NASA's first radio telescope on the lunar far side to be emplaced by a CLPS Firefly Aerospace lander in 2025/6 with the goal of making pathfinder observations in the frequency band corresponding to the global redshifted 21-cm signal for the Dark Ages (Bale et al. 2023).

In concert with these science missions are developing plans to demonstrate prototype in-situ extraction and manufacturing hardware and processes on the lunar surface this decade. With these new missions completed, we will be ready to create a high TRL concept for *FarView* and be ready to begin implementation during the 2030s.

**Acknowledgements**

The development of this lunar observatory concept was supported by the NASA Innovative Advanced Concepts (NIAC) program via grant 80NSSC21K0693 and internal LUNAR funding. This work was also directly supported by the NASA Solar System Exploration Virtual Institute cooperative agreement 80ARC017M0006. Part of this work was done at Jet Propulsion Laboratory, California Institute of Technology, under a contract with the National Aeronautics and Space Administration. We also wish to acknowledge the insightful and helpful comments from the reviewers of this paper – responding to their comments greatly improved the quality of the manuscript.

# Figures

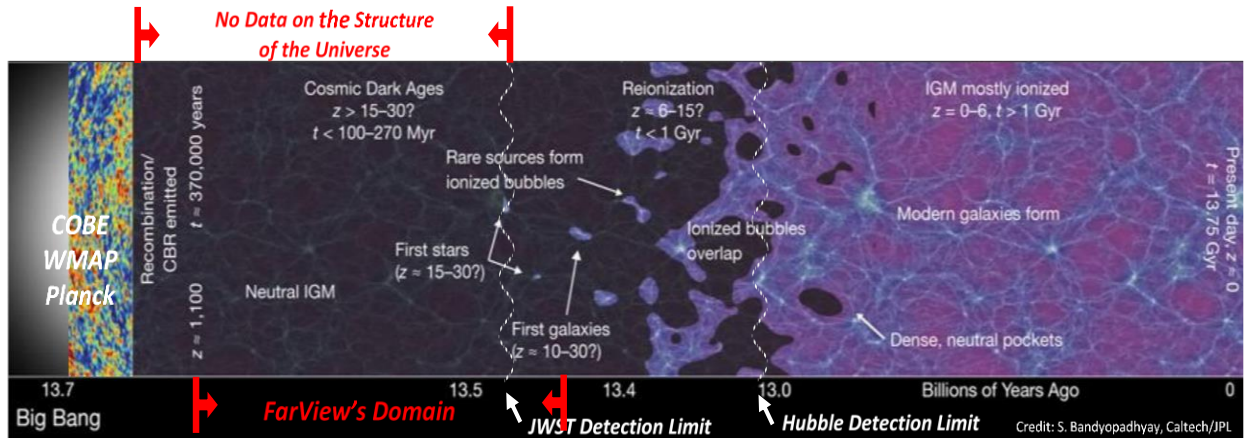

**Figure 1.** The pre-stellar (Dark Ages), first stars (Cosmic Dawn), and Reionization epochs of the Universe can be uniquely probed using the redshifted 21-cm signal. This history is accessible via the neutral hydrogen spin-flip background. *FarView* will fill in the missing data during the Cosmic Dark Ages and Cosmic Dawn. *Credit:* Adapted from JPL/Caltech (Bandyopadhyay et al. 2021).

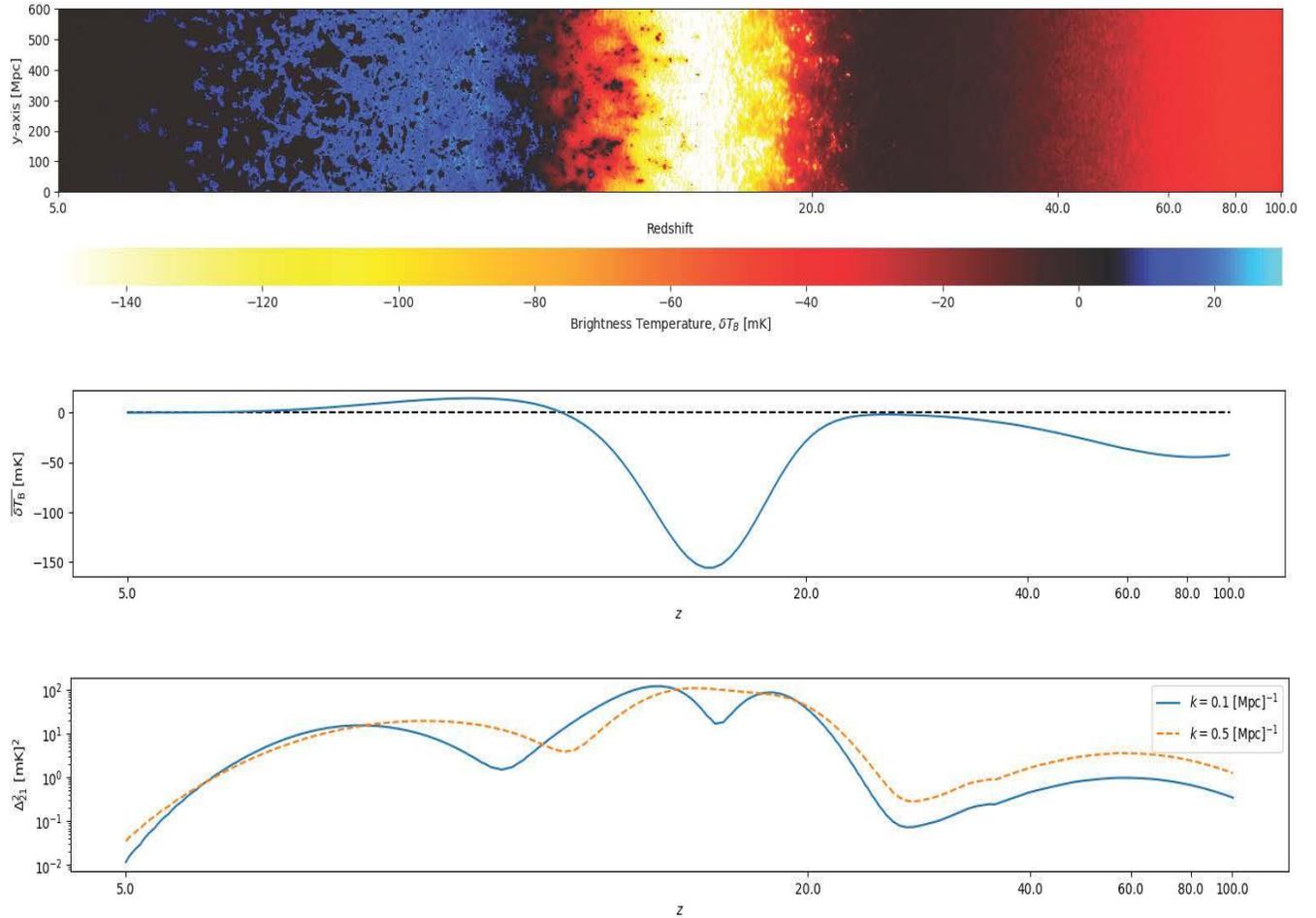

**Figure 2**: The redshifted 21-cm signal is a powerful probe of the evolution of the first structures in the Universe as well as providing tests of the standard models of physics and of cosmology. *Top:* A 2D slice through the "light cone" of the 21-cm brightness temperature field, illustrating the complex morphology of the signal and its potential for both emission and absorption relative to the CMB depending on the dominant coupling for the spin temperature of the hydrogen gas as a function of redshift. At z = 30 (end of Dark Ages, start of Cosmic Dawn), we expect a relatively weak absorption signal. *Middle:* The sky-averaged or "global" 21-cm signal versus redshift (blue line). The dashed black line runs along $\delta T_B = 0$ to guide the eye as to whether the 21-cm signal is in emission or absorption. *Bottom:* The amplitude of two Fourier modes ($k = 0.1$ Mpc$^{-1}$, blue solid, and $k = 0.5$ Mpc$^{-1}$, orange dashed, where the wavenumber $k = 2\pi/L$ Mpc$^{-1}$) of the dimensionless 21-cm power as a function of redshift generated using *21cmFast* for the standard ΛCDM cosmological model.

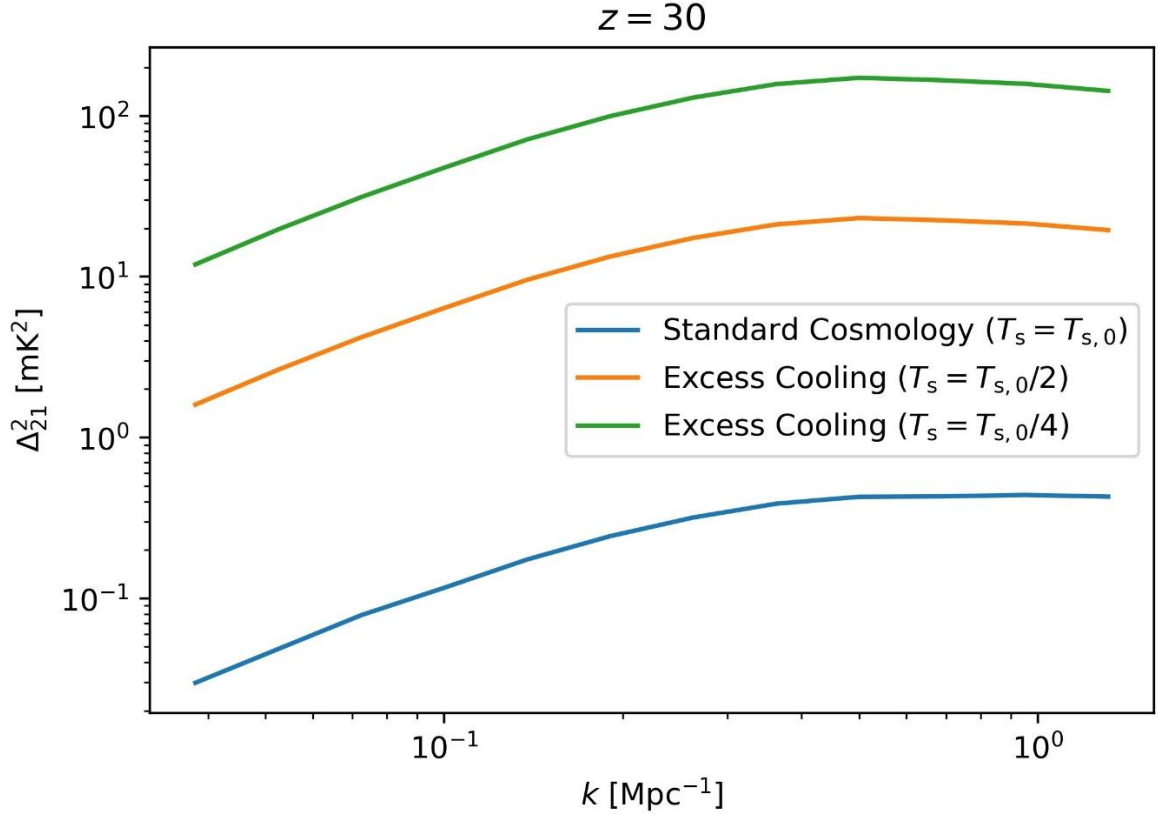

**Figure 3**. The 21-cm power spectrum can distinguish between different exotic physics scenarios during the Dark Ages. The dimensionless 21 cm power spectrum is shown here at z = 30. The standard cosmology (blue line) comes from using the default parameters of *21cmFAST* v3.3.1 (Mesinger, Furlanetto, & Cen 2011); the full evolution history of the 21-cm signal in this model is shown in Figure 2. The excess cooling models (orange and green lines) show the potential for enhancement of the 21 cm signal if the spin temperature of the neutral hydrogen is lower than that predicted by *21cmFAST*. Orange and green lines represent models where the spin temperature field has been scaled down by a factor of 2 and 4, respectively. Several mechanisms for this excess cooling have been suggested in the literature, including interactions between baryons and (possibly charged) dark matter (e.g., Barkana 2018, Muñoz & Loeb 2018).

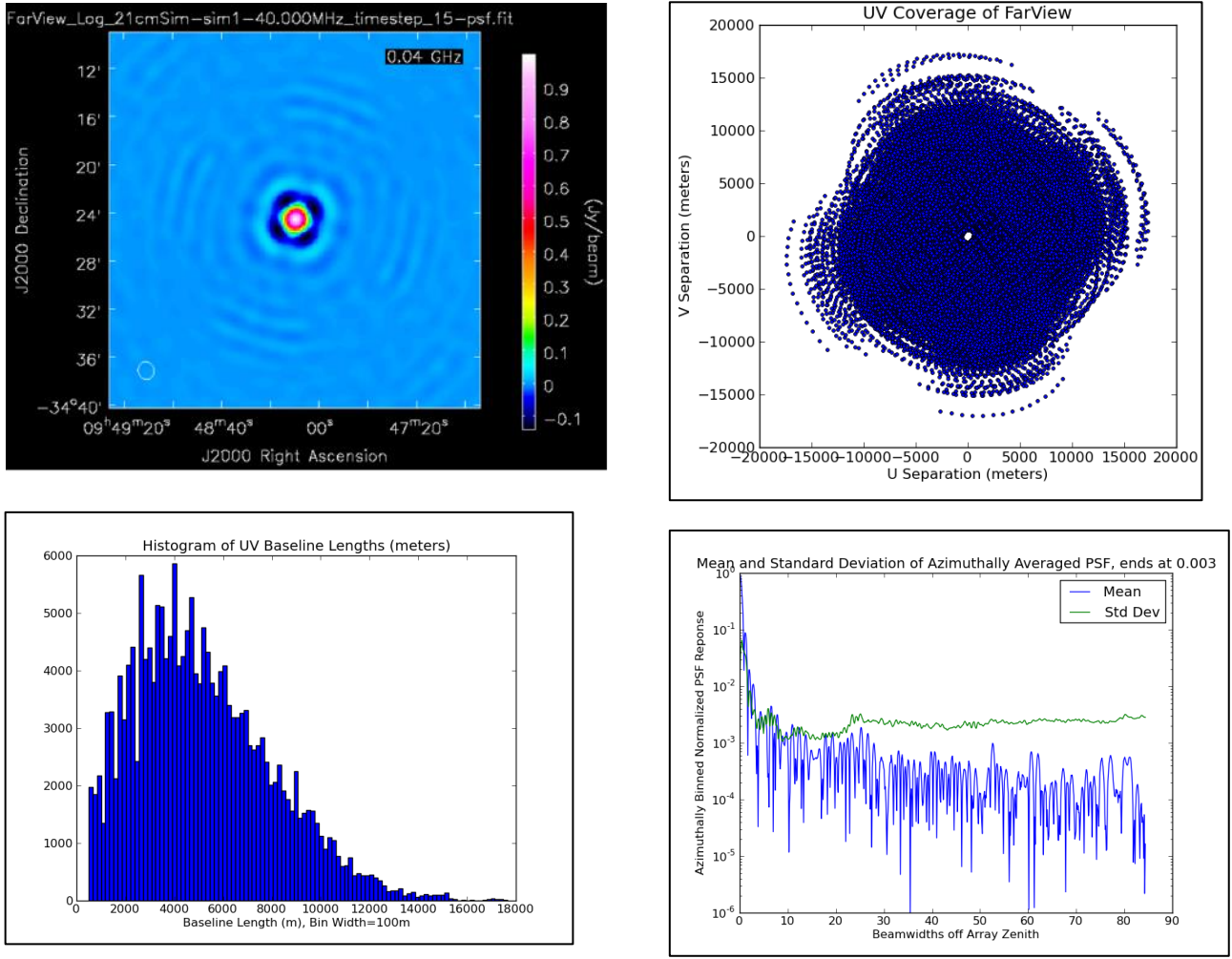

**Figure 4.** *Upper Left:* Point spread function (PSF) from *FarView*'s 160 subarrays spread shown in Figure 5. This simulated observation uses integration over 5 Earth days, 1-5 January 2035, over 15×8 hour discrete steps following the patch of sky that crosses zenith halfway through. The color scale here has been altered from linear to exaggerate small features. *Upper Right*: u-v coverage of this observation. The integration over time increases the u-v coverage from the relative rotation of the Moon and the sky, improving the response of the point spread function. *Bottom Left*: Histogram of u-v baseline lengths, showing a hole under 500 m, the minimum spacing between subarrays. *Bottom Right*: Azimuthally averaged PSF statistics, showing the mean and standard deviation for each radial bin off zenith. Smaller is better here, allowing more faithful reconstructions of the sky brightness patterns in the dirty images.

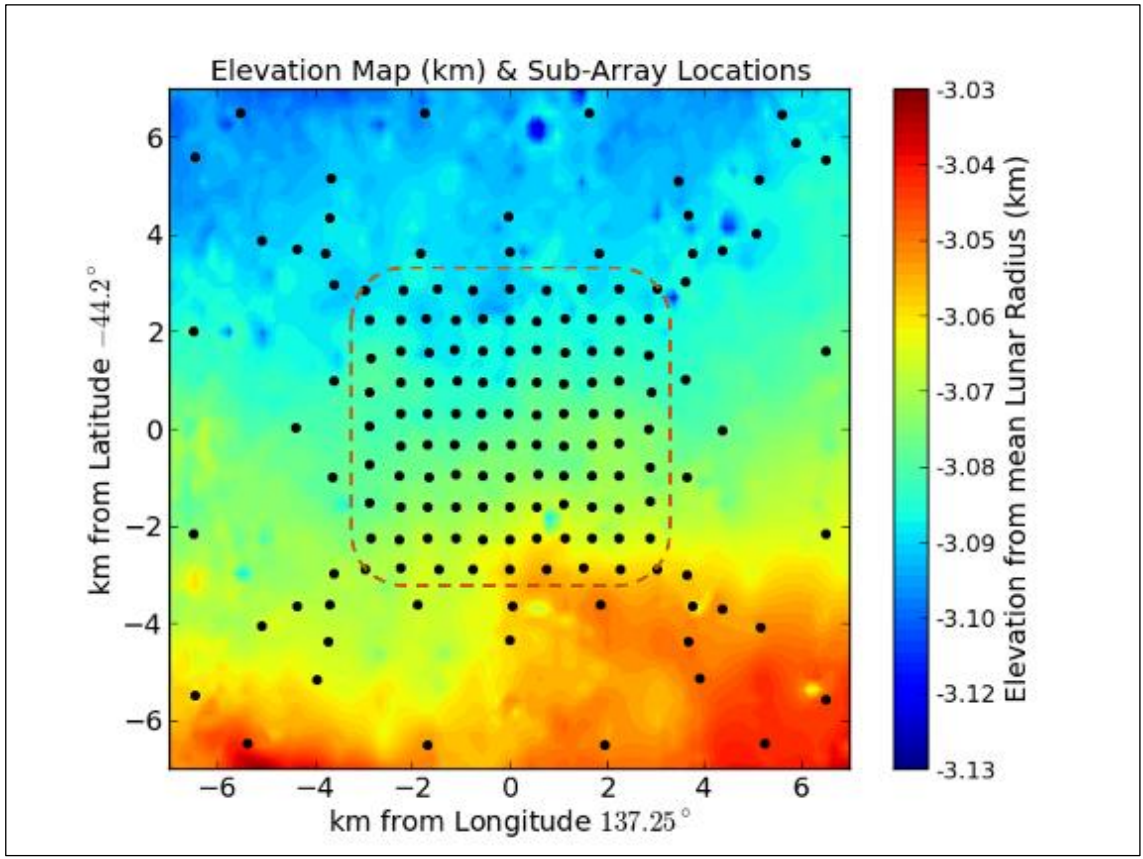

**Figure 5**. An approximate layout of subarrays for *FarView*. The array consists of a compact core (dashed line) with uniform spacings of antennas out to a diameter of 6 km and an outer power-law distribution of antennas out to a diameter of 14 km. This provides a combination of sensitivity to measure the Dark Ages power spectrum and good resolution to precisely measure the radio foreground. As an example, we show *FarView* sited within the Pauli impact basin on the lunar far side. This configuration is plotted over the Lunar DEMs from LRO/LOLA within the Pauli impact basin.

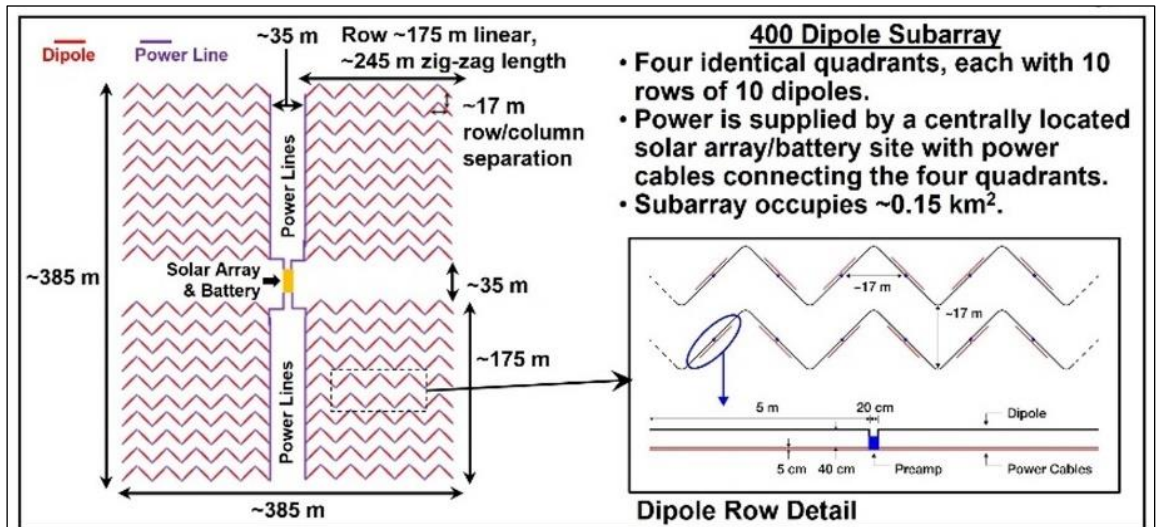

**Figure 6.** An illustrative example of a 400 dipole subarray showing dimensions and subarray details.

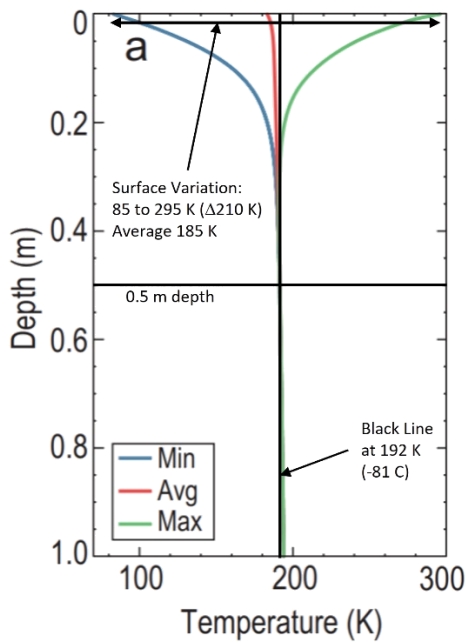

**Figure 7.** Temperature versus depth at the Cheng'E-4 landing site in Von Kármán crater. Figure is adapted from Figure 4 in Xiao et al. 2022.

# Tables

| Criteria | Weight | Sub-array | Score | Individual dipoles (no sub-arrays) | Score |
|---|---|---|---|---|---|
| **Computational requirement** | 25% | Reduces the computations required by reducing the u-v points/number of visibilities. | 5 | $N^2$ computational steps with N = 100,000 is large. Best case with EPIC is still Nlog(N). | 3 |
| **Deployment** | 10% | Easy rover operations | 4 | More distance & time travelled by rover due to larger number of cables | 3 |
| | 10% | Fewer cables (less mass and volume) | 5 | Cables from each dipole to the central node | 2 |
| | 10% | Fewer correlators on the surface | 4 | 100,000 correlators | 2 |
| **Heritage** | 5% | Building correlators for ~300-400 element array. Pipeline and simulation tools for 300-400 element array - heritage exists | 4 | This will be the largest array ever built | 3 |
| **Field-of-View (FOV)** | 5% | Reduces the instantaneous FOV from ~all-sky to $\lambda/(N_{sub}^{1/2} \times 17m) \sim 2°$ | 4 | All sky field of View | 5 |
| **Cadence of coverage** | 5% | Reduced | 4 | Maximum possible | 5 |
| **Blind transient detection rate** | 5% | Decreased | 4 | Best possible | 5 |
| **Sample/Cosmic variance** | 5% | Higher | 4 | Lower | 5 |
| **Foreground contamination** | 5% | Lower contamination | 5 | More contamination due to larger FoV | 4 |
| **Chromaticity** | 10% | Higher | 3 | Lower | 5 |
| **Operational Complexity** | 5% | Steerable beams | 3 | Drift scan | 5 |
| **Totals** | **100%** | | **4.1** | | **3.3** |

**Table 1.** Trade study for the operation of *FarView* using either beamforming of subarrays or full interferometry/correlation of all dipole antennas.

| Table 2 - Potential *FarView* Build Sites ||||
| Nomenclature | Longitude (degrees) | Latitude (degrees) | Type |
| --- | --- | --- | --- |
| **First Tier Locations** (Meet all the major selection criteria) ||||
| Pauli | 137.2 | -44.1 | Crater |
| Isaev | 147.9 | -17.5 | Crater |
| Crocco | 150.3 | -46.9 | Crater |
| Kohlschütter | 153.8 | 13.8 | Crater |
| van de Graaff | 173.6 | -25.3 | Crater |
| Lacus Luxuriae | 175.3 | 19.1 | Mare |
| Von Kármán | 177.53 | -45.9 | Crater |
| Unnamed Small Crater | 181.1 | 23.5 | Crater |
| **Second Tier Locations** (Meet all but one of the major selection criteria) ||||
| Tsiolkovsky[1] | 128.8 | -22.3 | Crater |
| Mare Muscoviense[1] | 151.6 | 27.7 | Mare |
| Thomson/Mare Ingenii[1] | 167.4 | -33.2 | Crater/Mare |
| Unnamed Crater SE of Von Kármán[2] | 171.9 | -51.0 | Crater |
| Leibnitz[1] | 179.1 | -35.8 | Crater |
| Daedalus[2] | 179.3 | -6.3 | Crater |
| Engelhardt B[2] | 201.2 | 7.6 | Crater |
| Southern Apollo[1] | 204.0 | -41.7 | Basin |

Selection Criteria Shortfall:
 1 Lower than desired aluminum abundance.
 2 Higher than desired crater abundance.